\documentclass[twocolumn,pre]{revtex4-1}
\usepackage{amsmath}
\usepackage{amssymb}
\usepackage{graphicx}
\usepackage{esint}
\begin{document}

\title{Thermodynamics of collective enhancement of precision}

\author{Yoshihiko Hasegawa}

\email{hasegawa@biom.t.u-tokyo.ac.jp}

\date{\today}

\affiliation{Department of Information and Communication Engineering, Graduate
School of Information Science and Technology, The University of Tokyo,
Tokyo 113-8656, Japan}
\begin{abstract}
The circadian oscillator exhibits remarkably high temporal precision,
despite its exposure to several fluctuations. The central mechanism
that protects the oscillator from fluctuations is a collective enhancement
of precision, where a population of coupled oscillators displays higher
temporal precision than that achieved without coupling. Since coupling
is essentially information exchange between oscillators, we herein
investigate the relation between the temporal precision and the information
flow between oscillators in the linearized Kuramoto model by using
stochastic thermodynamics. For general coupling, we find that the
temporal precision is bounded from below by the information flow.
We generalize the model to incorporate a time-delayed coupling and
demonstrate that the same relation also holds for the time-delayed
case. Furthermore, the temporal precision is demonstrated to be improved
in the presence of the time delay, and we  show that the increased
information flow is responsible for the time-delay-induced precision
improvement.

\end{abstract}
\maketitle

\section{Introduction}

The circadian oscillator is a biological clock that is prevalent in
biological organisms ranging from bacteria to humans \cite{Refinetti:2005:CircBook,Ukai:2010:MamCircadian}.
Since the circadian oscillator is induced by biochemical reactions
on a cellular level, it is subject to several fluctuations. Despite
such stochasticity, the circadian oscillator is known to exhibit incredibly
high temporal precision \cite{Moortgat:2000:Precision}, where the
temporal standard deviation is approximately $3$ to $5$ min in $24$
h \cite{Enright:1980:CircPrec}. In a single oscillator level, a phase
response curve, which quantifies the dynamics of the oscillator, is
optimized so as to achieve high temporal precision \cite{Hasegawa:2013:OptimalPRC,Hasegawa:2014:PRL}.
Still, this single oscillator level improvement does not seem to fully
explain the abovementioned temporal precision. Another significant
precision improvement arises at the population level \cite{Winfree:2001:GeoBiolTime}.
A previous study \cite{Clay:1979:CEP} experimentally showed that
when oscillators are coupled, each oscillator exhibits higher precision
than that realizable without coupling. This phenomenon is referred
to as \emph{collective enhancement of precision} (CEP) \cite{Winfree:2001:GeoBiolTime,Needleman:2001:CEP,Kori:2012:OscReg}.
Since mammalian circadian oscillation occurs primarily in the suprachiasmatic
nucleus (SCN), which is a collection of $10^{4}$ cells, CEP is considered
to be responsible for the high temporal precision in the circadian
oscillator \cite{Winfree:2001:GeoBiolTime}.

\begin{figure}
\includegraphics[width=8cm]{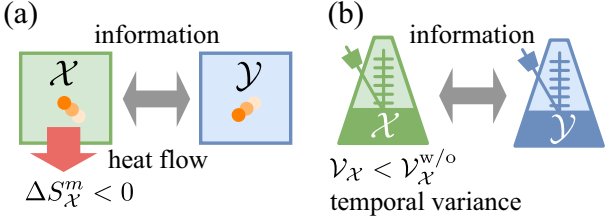}

\protect\caption{Analogy between (a) the Maxwell demon and (b) CEP. (a) When a system
$\mathcal{X}$ is coupled to another system $\mathcal{Y}$, $\mathcal{X}$'s
medium entropy $\Delta S_{\mathcal{X}}^{m}$, which is the heat from
$\mathcal{X}$ to the medium (divided by the temperature), can be
negative. (b) When an oscillator $\mathcal{X}$ is coupled to another
oscillator $\mathcal{Y}$, $\mathcal{V}_{\mathcal{X}}$ (the temporal
variance with coupling) can be smaller than $\mathcal{V}_{\mathcal{X}}^{\mathrm{w/o}}$(the
temporal variance without coupling). \label{fig:analogy}}

\end{figure}

CEP is induced by coupling, which is essentially the exchange of
information. Information enables the violation of the second law of
thermodynamics through the Maxwell demon. High precision is achieved
at the cost of energy consumption, which is known as the thermodynamic
uncertainty relation \cite{Cao:2015:ClockEnergy,Barato:2015:UncRel,Barato:2016:BrClo,Gingrich:2016:TUP,Horowitz:2017:TUR}.
When we ignore the coupling between oscillators, CEP appears to achieve
thermodynamically impossible precision, motivating us to analyze CEP
from the viewpoint of information analogous to the Maxwell demon (Fig.~\ref{fig:analogy}).
Using stochastic thermodynamics \cite{Ritort:NEArticle,Seifert:2012:FTReview,VandenBroeck:2015:Review},
we find that the temporal variance of coupled oscillators is bounded
from below by the information flow conferred by coupling. Furthermore,
we generalize the obtained relation to incorporate time-delayed coupling
and show that the same relation also holds for the time-delayed case.
Intriguingly, the temporal precision in the time-delayed case is improved
as compared to the non-delayed case, and we can ascribe this time
delay induced improvement to the increase in the information flow.

\section{Model}

\begin{figure}
\includegraphics[width=8cm]{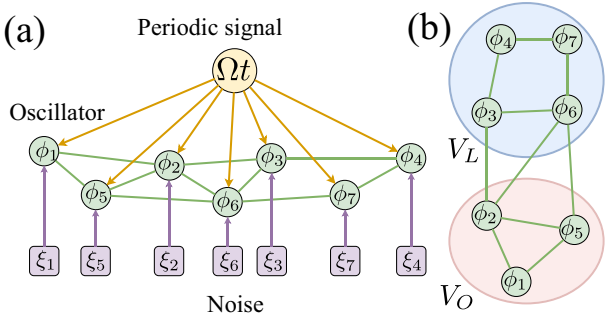}

\protect\caption{Coupled oscillators model. (a) Adopted coupled oscillator model for
$N=7$ ($V=\{1,2,...,7\}$). Each oscillator is entrained by a periodic
signal with the angular frequency $\Omega$ and is subject to noise
$\xi_{i}(t)$. (b) Observable set $V_{O}$ and latent set $V_{L}$,
where $V_{O}=\{1,2,5\}$ and $V_{L}=\{3,4,6,7\}$ in this example.
\label{fig:network_image}}
\end{figure}

We consider $N$ coupled identical oscillators, which are entrained
by an external periodic signal with the angular frequency $\Omega$
and are subject to noise (Fig.~\ref{fig:network_image}(a)). Let
$\phi_{i}$ be the phase of the $i$th oscillator, and let $\omega$
be the angular frequency of oscillators. Then, the phase dynamics
of each oscillator can be described by the following forced Kuramoto
model \cite{Sakaguchi:1988:KuramotoExt,Needleman:2001:CEP,Acebron:2005:KuramotoReview}:
\begin{align}
\dot{\phi}_{i} & =\omega+L\sin(\Omega t-\phi_{i})+f_{i}(t)+\xi_{i}(t),\label{eq:Kuramoto_model_def}\\
f_{i}(t) & =\sum_{j\in V,j\ne i}K_{ij}\sin(\phi_{j}-\phi_{i}),\label{eq:F_def_Kuramoto}
\end{align}
where $V=\{1,2,...,N\}$, and we aggregate all coupling effects, which
act on $\phi_{i}$, into a single coupling variable $f_{i}(t)$. The
sinusoidal term in Eq.~\eqref{eq:Kuramoto_model_def} represents
the periodic signal, and $L>0$ denotes its strength. A sinusoidal
function in Eq.~\eqref{eq:F_def_Kuramoto} is a coupling function
that represents the interaction between oscillators, where $K_{ij}$
is the coupling strength between the $i$th and $j$th oscillators.
Note that we allow both symmetric ($K_{ij}=K_{ji}$) and asymmetric
($K_{ij}\ne K_{ji}$) couplings. The phase reduction, with which the
phase equation of Eq.~\eqref{eq:Kuramoto_model_def} is derived,
assumes Langevin equations to be interpreted in the Stratonovich sense.
Moreover, $\xi_{i}(t)$ is zero-mean white Gaussian noise with the
correlation $\left\langle \xi_{i}(t)\xi_{i'}(t')\right\rangle =2D\delta_{ii'}\delta(t-t')$
($D$ is the noise strength). The derivation of Eq.~\eqref{eq:Kuramoto_model_def}
can be found in Ref.~\cite{Pikovsky:2001:SyncBook,Kuramoto:2003:OscBook}.
We hereinafter assume that all of the oscillators are synchronized
to the periodic signal. Although CEP is often studied for $L=0$,
we assume $L>0$, which enables an approximation of the synchronized
behavior as a nonequilibrium steady state (NESS). 

Since there are efferent projections from the SCN to other neurons,
the output of the SCN has been suggested to be generated by the average
of a subset of oscillators, and not by the average of all oscillators
\cite{Kori:2012:OscReg}. Therefore, we observe the average phase
$\Phi(t)=(1/N_{O})\sum_{i\in V_{O}}\phi_{i}(t)$, where $V_{O}$ is
a set of observable oscillators, and $N_{O}$ is its element number
(Fig.~\ref{fig:network_image}(b)) \cite{Kori:2012:OscReg}. We also
define $V_{L}$ and $N_{L}$ to be a set of latent variables (the
output of which is not observed) and its element number, respectively.
Every oscillator belongs to the whole set $V=V_{O}\cup V_{L}$, and
$V_{O}$ and $V_{L}$ are disjoint, i.e., $V_{O}\cap V_{L}=\varnothing$.
See Fig.~\ref{fig:network_image}(b). Defining a relative phase $x_{i}$
observed from a rotating frame as $x_{i}=\phi_{i}-\Omega t,$ the
synchronizing assumption is that $x_{i}$ is well concentrated around
$0$. This condition can usually be satisfied when $|\omega-\Omega|/L$
is sufficiently small and $D$ is sufficiently weak. Let $\mu_{A}=\left\langle A\right\rangle $,
$\mathcal{V}_{A}=\left\langle A^{2}\right\rangle -\left\langle A\right\rangle ^{2}$,
and $\mathcal{C}_{AB}=\left\langle AB\right\rangle -\left\langle A\right\rangle \left\langle B\right\rangle $,
where $A$ and $B$ are arbitrary random variables, and let $X(t)=\Phi(t)-\Omega t=(1/N_{O})\sum_{i\in V_{O}}x_{i}(t)$,
which is the average phase seen from a rotating frame. The temporal
variance of the average phase $\Phi(t)$ is given by the variance
$\mathcal{V}_{X}$. Since each phase $x_{i}$ is well concentrated
around $0$ (we assumed that all of the oscillators are synchronized
with the periodic signal), we apply a Taylor series expansion to Eqs.~\eqref{eq:Kuramoto_model_def}
and \eqref{eq:F_def_Kuramoto} to obtain 
\begin{align}
\dot{X} & =-L(X-c)+F(t)+\Xi_{O}(t),\label{eq:dXdt_def_1}\\
\dot{x}_{i} & =-L(x_{i}-c)+\sum_{j\in V,j\ne i}K_{ij}(x_{j}-x_{i})+\xi_{i}(t)\hspace*{1em}(i\in V),\label{eq:dxdt_def_1}
\end{align}
where $F(t)$ is defined by 
\begin{equation}
F(t)=\frac{1}{N_{O}}\sum_{i\in V_{O}}f_{i}(t)=\frac{1}{N_{O}}\sum_{i\in V_{O}}\sum_{j\in V,j\ne i}K_{ij}(x_{j}-x_{i}),\label{eq:F_def}
\end{equation}
and $c=(\omega-\Omega)/L$ and $\Xi_{O}(t)=(1/N_{O})\sum_{i\in V_{O}}\xi_{i}(t)$
{[}$\left\langle \Xi_{O}(t)\Xi_{O}(t')\right\rangle =(2D/N_{O})\delta(t-t')$,
$\left\langle \Xi_{O}(t)\xi_{i}(t')\right\rangle =(2D/N_{O})\delta(t-t')$
for $i\in V_{O}$, and $\left\langle \Xi_{O}(t)\xi_{i}(t')\right\rangle =0$
for $i\in V_{L}${]}. Here, $F(t)$ includes all of the coupling effects
that act on $x_{i}$ for $i\in V_{O}$. Therefore, the information
flow between $X$ and $F$ is of interest. We calculate the information
flow between $X$ and $F$ based on Refs.~\cite{Horowitz:2014:InfoFlow,Horowitz:2015:Multipartite},
which is defined as
\begin{align}
\dot{\mathcal{I}}_{X}(X;F) & =-\int J_{X}(X,F)\frac{\partial}{\partial X}\ln P(F|X)dXdF\nonumber \\
 & =-\frac{1}{dt}\left\langle \ln\left(\frac{P(F(t)|X(t+dt))}{P(F(t)|X(t))}\right)\right\rangle ,\label{eq:I_X_def}
\end{align}
where the sign is reversed compared to the original definition. This
quantity is equivalent to the learning rate \cite{Barato:2014:CellInfo,Hartich:2016:SensCap,Brittain:2017:LR}.
In order to calculate the information flow, we introduce a hypothetical
time delay $h>0$ in Eq.~\eqref{eq:F_def}, which yields $F(t)=(1/N_{O})\sum_{i\in V_{O}}\sum_{j\in V,j\ne i}K_{ij}(x_{j}(t-h)-x_{i}(t-h))$.
The introduction of time delay $h$ is a mathematical trick, and,
when calculating several quantities, we set $h\rightarrow0$ afterwards.
Using a Taylor expansion for a sufficiently small $h$, we obtain
\begin{equation}
\dot{F}=h^{-1}\left(-F+\frac{1}{N_{O}}\sum_{i\in V_{O}}\sum_{j\in V,j\ne i}K_{ij}(x_{j}-x_{i})\right).\label{eq:dFdt_def}
\end{equation}
Then, we can treat Eqs.~\eqref{eq:dXdt_def_1}, \eqref{eq:dxdt_def_1},
and \eqref{eq:dFdt_def} as coupled Langevin equations. Let $P(X,\boldsymbol{x},F)$
be a time-dependent probability density function of $X$, $\boldsymbol{x}=\left[x_{1},...,x_{N}\right]$,
and $F$ (for notational convenience, we dropped time $t$ in the
argument in $P(X,\boldsymbol{x},F)$). Due to the linearity of Eqs.~\eqref{eq:dXdt_def_1},
\eqref{eq:dxdt_def_1}, and \eqref{eq:dFdt_def}, the steady-state
distribution is a Gaussian distribution with mean $\left[\mu_{X},\mu_{x_{1}},..,\mu_{x_{N}},\mu_{F}\right]=\left[c,c,...,c,0\right]$.
The corresponding Fokker--Planck equation (FPE) with respect to $P(X,\boldsymbol{x},F)$
can be represented as 
\begin{equation}
\partial_{t}P(X,\boldsymbol{x},F)=\mathbb{L}(X,\boldsymbol{x},F)P(X,\boldsymbol{x},F),\label{eq:FPE_XFx}
\end{equation}
where $\mathbb{L}(X,\boldsymbol{x},F)$ is an FPE operator, which
is defined by \begin{widetext} 
\begin{align}
\mathbb{L}(X,\boldsymbol{x},F) & =-\frac{\partial}{\partial X}\left(-L(X-c)+F\right)+\frac{D}{N_{O}}\frac{\partial^{2}}{\partial X^{2}}-\frac{\partial}{\partial F}\frac{1}{h}\left\{ -F+\frac{1}{N_{O}}\sum_{i\in V_{O}}\sum_{j\in V,j\ne i}K_{ij}\left(x_{j}-x_{i}\right)\right\} \nonumber \\
 & -\sum_{i\in V}\frac{\partial}{\partial x_{i}}\left\{ -L(x_{i}-c)+\sum_{j\in V,j\ne i}K_{ij}\left(x_{j}-x_{i}\right)\right\} +D\sum_{i\in V}\frac{\partial^{2}}{\partial x_{i}^{2}}+\frac{2D}{N_{O}}\sum_{i\in V_{O}}\frac{\partial^{2}}{\partial x_{i}\partial X}.\label{eq:FPE_XFx_1-1}
\end{align}
\end{widetext}We integrate out $\boldsymbol{x}$ in Eqs.~\eqref{eq:FPE_XFx}
and \eqref{eq:FPE_XFx_1-1} to obtain 
\begin{align}
\partial_{t}P(X,F) & =-\partial_{X}J_{X}(X,F)-\partial_{F}J_{F}(X,F),\label{eq:FPE_XF}
\end{align}
where $J_{X}(X,F)$ and $J_{F}(X,F)$ are probability currents, which
are defined as 
\begin{align}
J_{X}(X,F) & =\left\{ -L(X-c)+F\right\} P-\frac{D}{N_{O}}\frac{\partial P}{\partial X},\label{eq:JX_def}\\
J_{F}(X,F) & =h^{-1}\left\{ -F+\mathcal{H}(X,F)\right\} P.\label{eq:JF_def}
\end{align}
in which $P=P(X,F)$ and $\mathcal{H}(X,F)=\int(1/N_{O})\sum_{i\in V_{O}}\sum_{j\in V,j\ne i}K_{ij}\left(x_{j}-x_{i}\right)P(\boldsymbol{x}|X,F)d\boldsymbol{x}$.
Although the FPE operator $\mathbb{L}(X,\boldsymbol{x},F)$ of Eq.~\eqref{eq:FPE_XFx_1-1}
contains cross terms, such as $\partial_{X}\partial_{x_{i}}$, due
to non-vanishing correlation $\left\langle \Xi_{O}(t)\xi_{i}(t)\right\rangle \ne0$
for $i\in V_{O}$, they disappear when integrating out $\boldsymbol{x}$.
From Eq.~\eqref{eq:FPE_XF}, we can obtain the generalized second
law following the conventional procedure in stochastic thermodynamics
\cite{Seifert:2012:FTReview,Horowitz:2014:InfoFlow,VandenBroeck:2015:Review,Horowitz:2015:Multipartite}.
We first consider the Shannon entropy $S(X,F)=-\int P(X,F)\ln P(X,F)dXdF$.
The time derivative is $\dot{S}(X,F)=\dot{S}_{X}(X,F)+\dot{S}_{F}(X,F)$,
where $\dot{S}_{X}(X,F)=-\int J_{X}(X,F)\partial_{X}\ln P(X,F)dXdF$,
and $\dot{S}_{F}(X,F)=-\int J_{F}(X,F)\partial_{F}\ln P(X,F)dXdF$.
We can divide $J_{X}(X,F)$ of Eq.~\eqref{eq:JX_def} into two contributions,
$J_{X}^{R}(X,F)$ and $J_{X}^{I}(X,F)$, which are defined as $J_{X}^{I}(X,F)=-L(X-c)P-(D/N_{O})\partial_{X}P$
and $J_{X}^{R}=FP$, respectively. In feedback cooling of Brownian
particles, $J_{X}^{I}$ and $J_{X}^{R}$ correspond to the irreversible
and reversible portions of the probability current \cite{Munakata:2013:Feedback,Horowitz:2014:InfoFlow}.
Then, $\dot{S}_{X}(X,F)$ is calculated as 
\begin{equation}
\dot{S}_{X}(X,F)=\int\left(\frac{N_{O}\left(J_{X}^{I}\right)^{2}}{DP}+\frac{N_{O}J_{X}^{I}L(X-c)}{D}\right)dXdF,\label{eq:SX_dot_XF}
\end{equation}
where $J_{X}^{I}=J_{X}^{I}(X,F)$, and $P=P(X,F)$. In Eq.~\eqref{eq:SX_dot_XF},
the first term is non-negative, and the second term is calculated
as $(1/D)\int N_{O}J_{X}^{I}(X,F)L(X-c)dXdF=-\left(N_{O}L^{2}/D\right)\left\langle (X-c)^{2}\right\rangle +L$.
According to Ref.~\cite{Horowitz:2015:Multipartite}, we have $\dot{\mathcal{I}}_{X}(X;F)=-\dot{S}(X)+\dot{S}_{X}(X,F)$,
where $S(X)=-\int P(X)\ln P(X)dX$. Under the synchronization assumption,
$X$ is in a steady state, and thus $\dot{S}(X)=0$. Combining these
representations, the variance $\mathcal{V}_{X}=\left\langle (X-c)^{2}\right\rangle $
satisfies 
\begin{equation}
\mathcal{V}_{X}\ge\mathcal{V}_{X}^{\mathrm{LB}}=\frac{D}{L^{2}N_{O}}\left(L-\dot{\mathcal{I}}_{X}(X;F)\right),\label{eq:LB_def}
\end{equation}
where $\mathcal{V}_{X}^{\mathrm{LB}}$ is the lower bound of the variance
$\mathcal{V}_{X}$. Equation~\eqref{eq:LB_def} is our main result.
When there is no coupling, $\mathcal{V}_{X}=D/(LN_{O})$. Therefore,
without information flow, the variance does not improve beyond $D/(LN_{O})$,
and Eq.~\eqref{eq:LB_def} shows that CEP is induced by the information
flow due to coupling between oscillators. Equation~\eqref{eq:LB_def}
indicates that CEP has the same mathematical property as the feedback
cooling by the Maxwell demon \cite{Horowitz:2014:SecLawLike,Ito:2015:Maxwell},
where the feedback reduces the variance. The angular frequency $\Omega$
affects the inequality of Eq.~\eqref{eq:LB_def} through $c=(\omega-\Omega)/L$,
showing that the inequality does not have $\Omega$ dependence when
$\omega=\Omega$.  Note that we cannot reach the desired inequality
of Eq.~\eqref{eq:LB_def} by using the positivity of $\int J_{X}(X,F)^{2}/P(X,F)dXdF$
in the conventional total entropy production rate \cite{Tome:2006:EPinFPE}.
The transfer entropy rate \cite{Schreiber:2000:TI} is a similar
quantity to the information flow. It has been shown that the transfer
entropy rate from $X$ to $F$ is greater than or equal to the information
flow \cite{Allahverdyan:2009:TE,Hartich:2014:Bipartite,Horowitz:2015:Multipartite}.
Therefore, the same inequality as Eq.~\eqref{eq:LB_def} holds for
the transfer entropy rate but the bound becomes weaker than the information
flow case.

\section{Examples}

\subsection{Globally coupled model}

\begin{figure}
\includegraphics[width=8.5cm]{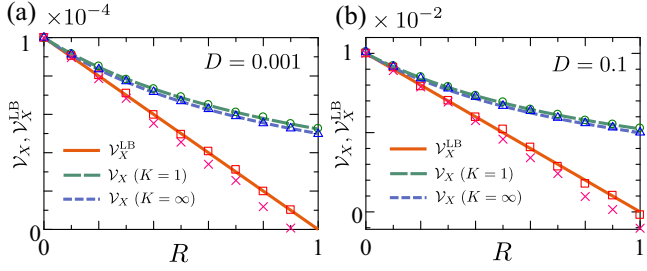}

\protect\caption{Variance $\mathcal{V}_{X}$ and its lower bound $\mathcal{V}_{X}^{\mathrm{LB}}$
as a function of $R=N_{L}/N_{O}$ for (a) $D=0.001$ and (b) $D=0.1$.
The solid, long-dashed, and dashed lines denote $\mathcal{V}_{X}^{\mathrm{LB}}$,
$\mathcal{V}_{X}$ with $K=1$, and $\mathcal{V}_{X}$ with $K=\infty$,
respectively, obtained analytically. The squares, crosses, circles,
and triangles denote $\mathcal{V}_{X}^{\mathrm{LB}}$ with $K=1$,
$\mathcal{V}_{X}^{\mathrm{LB}}$ with $K=10$, $\mathcal{V}_{X}$
with $K=1$, and $\mathcal{V}_{X}$ with $K=10$, respectively, obtained
by Monte Carlo simulation. The other parameters are $N_{O}=10$, $L=1$,
$\omega=1$, and $\Omega=1$. \label{fig:variance_plot}}
\end{figure}

We calculate $\mathcal{V}_{X}$ and $\mathcal{V}_{X}^{\mathrm{LB}}$
analytically for a global uniform coupling case $K_{ij}=K$. Defining
$Y=(1/N_{L})\sum_{i\in V_{L}}x_{i}$, we have 
\begin{align}
\dot{Y} & =-L(Y-c)-F/R+\Xi_{L}(t),\label{eq:dYdt_def}\\
\dot{F} & =h^{-1}\left\{ -F+KN_{L}(Y-X)\right\} ,\label{eq:dFdt_gl_def}
\end{align}
where $R=N_{L}/N_{O}$ and $\Xi_{L}(t)=(1/N_{L})\sum_{i\in V_{L}}\xi_{i}(t)$
{[}$\left\langle \Xi_{L}(t)\Xi_{L}(t')\right\rangle =(2D/N_{L})\delta(t-t')$
and $\left\langle \Xi_{L}(t)\Xi_{O}(t')\right\rangle =0${]}. Then,
Eqs.~\eqref{eq:dXdt_def_1}, \eqref{eq:dYdt_def}, and \eqref{eq:dFdt_gl_def}
constitute coupled Langevin equations of the global uniform coupling
model. Taking a limit $h\rightarrow0$ for which the hypothetical
time delay in $F$ vanishes, we can calculate $\mathcal{V}_{X}$,
$\mathcal{V}_{F}$, and $\mathcal{C}_{XF}$ analytically due to the
linearity of the coupled equations. Then, the information flow $\dot{\mathcal{I}}_{X}(X;F)$
can be calculated as follows (Appendix~\ref{sec:app_glob_couple}):
\begin{equation}
\dot{\mathcal{I}}_{X}(X;F)=LR.\label{eq:IX_global}
\end{equation}
We plot $\mathcal{V}_{X}^{\mathrm{LB}}$ and $\mathcal{V}_{X}$ of
the global uniform coupling case for $D=0.001$ (Fig.~\ref{fig:variance_plot}(a))
and $D=0.1$ (Fig.~\ref{fig:variance_plot}(b)). The other parameters
are shown in the caption of Fig.~\ref{fig:variance_plot}. In Fig.~\ref{fig:variance_plot},
the solid line denotes $\mathcal{V}_{X}^{\mathrm{LB}}$, and the long-dashed
and dashed lines show $\mathcal{V}_{X}$ for $K=1$ and $K=\infty$,
respectively. We also carried out a Monte Carlo simulation (for details,
please see Appendix~\ref{sec:app_monte_carlo}), where $\mathcal{V}_{X}$
are denoted by circles ($K=1$) and triangles ($K=10$, which is intended
to simulate $K=\infty$), and $\mathcal{V}_{X}^{\mathrm{LB}}$ are
denoted by squares ($K=1$) and crosses ($K=10$). Note that the Monte
Carlo simulation is carried out for equations that do not use the
Taylor approximation. Apparently, $\mathcal{V}_{X}\ge\mathcal{V}_{X}^{\mathrm{LB}}$
for $R\ge0$, which verifies the inequality relation. For $R>1$,
$\mathcal{V}_{X}^{\mathrm{LB}}$ becomes negative, and the lower bound
does not have a practical meaning. The difference is $\mathcal{V}_{X}-\mathcal{V}_{X}^{\mathrm{LB}}=L^{-2}\int J_{X}^{I}(X,F)^{2}/P(X,F)dXdF$
and the probability current $|J_{X}^{I}|$ becomes larger for larger
$R$, which is responsible for the gap between $\mathcal{V}_{X}^{\mathrm{LB}}$
and $\mathcal{V}_{X}$. The inequality saturates when $F\rightarrow0$,
i.e., there is no coupling. Figure~\ref{fig:variance_plot} shows
that the Monte Carlo results agree with the analytical calculations,
even for a relatively large noise intensity of $D=0.1$ (Fig.~\ref{fig:variance_plot}(b)),
and this demonstrates the validity of the synchronization assumption
and the Taylor approximation. When $R$ is sufficiently small, the
variance $\mathcal{V}_{X}$ is close to $\mathcal{V}_{X}^{\mathrm{LB}}$,
which shows that oscillators maximally exploit the information flow
to improve the variance. As shown in the next model, higher temporal
precision can be achieved if the information flow $\dot{\mathcal{I}}_{X}(X;F)$
can be increased.

\begin{figure}
\includegraphics[width=8.5cm]{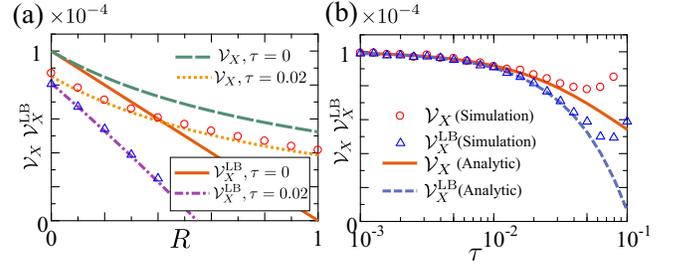}

\protect\caption{Variance $\mathcal{V}_{X}$ and its lower bound $\mathcal{V}_{X}^{\mathrm{LB}}$for
a time-delayed case. (a) $\mathcal{V}_{X}$ and $\mathcal{V}_{X}^{\mathrm{LB}}$
as a function of $R=N_{L}/N_{O}$ for the time-delayed case ($\tau=0.02$)
and $K=1$. The dotted and dot-dashed lines denote $\mathcal{V}_{X}$
and $\mathcal{V}_{X}^{\mathrm{LB}}$ for $\tau=0.02$, respectively,
obtained analytically. The solid and long-dashed lines denote $\mathcal{V}_{X}^{\mathrm{LB}}$
and $\mathcal{V}_{X}$, respectively, for the non-delayed case ($\tau=0$)
and are identical to those in Fig.~\ref{fig:variance_plot}(a). The
circles and triangles denote $\mathcal{V}_{X}$ and $\mathcal{V}_{X}^{\mathrm{LB}}$,
respectively, for $\tau=0.02$ obtained by Monte Carlo simulations.
(b) Variance $\mathcal{V}_{X}$ and its lower bound $\mathcal{V}_{X}^{\mathrm{LB}}$
as a function of the time delay $\tau$ for $K=1$ and $N=N_{O}=10$.
The solid and dashed lines denote $\mathcal{V}_{X}$ and $\mathcal{V}_{X}^{\mathrm{LB}}$,
respectively, obtained analytically. The circles and triangles denote
$\mathcal{V}_{X}$ and $\mathcal{V}_{X}^{\mathrm{LB}}$, respectively,
obtained by Monte Carlo simulations. In (a) and (b), the other unspecified
parameters are the same as those in Fig.~\ref{fig:variance_plot}(a).
\label{fig:delayed_fig}}
\end{figure}

\subsection{Time-delayed model}

Next, we generalize the model to include time delay in the coupling,
which exists in reality and has been reported to be able to cause
different dynamics in the Kuramoto model \cite{Yeung:1999:DelayedKuramoto}.
In the context of feedback cooling, a time-delayed case was investigated
in Ref.~\cite{Rosinberg:2015:Delay,Rosinberg:2017:Delay}. Let $\tau\ge0$
be the time delay of the coupling. Following Ref.~\cite{Yeung:1999:DelayedKuramoto},
we incorporate the time delay into Eq.~\eqref{eq:Kuramoto_model_def}
using the following $f_{i}(t)$ instead of Eq.~\eqref{eq:F_def_Kuramoto}:
\begin{equation}
f_{i}(t)=\sum_{j\in V,j\ne i}K_{ij}\sin(\phi_{j}^{\tau}-\phi_{i}),\label{eq:fi_def_delay}
\end{equation}
where superscript $\tau$ hereinafter denotes a time-delayed variable,
i.e., $\phi_{j}^{\tau}(t)=\phi_{j}(t-\tau)$. Similar to the non-delayed
case, we introduce a relative phase $x_{i}=\phi_{i}-\Omega t$. For
the non-delayed case, $\mu_{X}=0$ when $\omega=\Omega$. However,
for the time-delayed case, $\mu_{X}$ deviates from $0$, even when
$\omega=\Omega$. Therefore, in order for the Taylor approximation
to yield reliable results (i.e., $x_{i}$ should be well concentrated
around $0$), $\Omega\tau$ should be sufficiently small. In such
cases, we can use the Taylor approximation to obtain $F=(1/N_{O})\sum_{i\in V_{O}}f_{i}=(1/N_{O})\sum_{i\in V_{O}}\sum_{j\in V,j\ne i}K_{ij}(x_{j}^{\tau}-x_{i}-\Omega\tau)$,
and the introduction of the hypothetical delay $h$ leads us to $\dot{F}=(1/h)\left\{ -F+(1/N_{O})\sum_{i\in V_{O}}\sum_{j\in V,j\ne i}K_{ij}(x_{j}^{\tau}-x_{i}-\Omega\tau)\right\} $.
Following Ref.~\cite{Guillouzic:1999:SDDE}, an effective FPE, which
does not include time-delayed variables, can be derived for the time-delayed
case. Let $\boldsymbol{x}^{\tau}=[x_{1}^{\tau},...,x_{N}^{\tau}]$
and $P(X,\boldsymbol{x},F;\boldsymbol{x}^{\tau})$ be the probability
density of $X$, $\boldsymbol{x}$, and $F$ at time $t$, and $\boldsymbol{x}^{\tau}$
at time $t-\tau$. Calculating the marginal distribution $P(X,F)=\int P(X,\boldsymbol{x},F;\boldsymbol{x}^{\tau})d\boldsymbol{x}d\boldsymbol{x}^{\tau}$,
the FPE for $P(X,F)$ is similar to that given by Eqs.~\eqref{eq:FPE_XF}--\eqref{eq:JF_def}.
Therefore, the inequality as in Eq.~\eqref{eq:LB_def} holds for
the time-delayed case (Appendix~\ref{sec:app_time_delay}).

\subsection{Globally coupled time-delayed model}

Next, we calculate $\mathcal{V}_{X}$ and $\mathcal{V}_{X}^{\mathrm{LB}}$
for a globally coupled time-delayed case. For this case, we obtained
the following equations: 
\begin{align}
\dot{X} & =-LX+F+\Xi_{O}(t),\label{eq:dXdt_GTD_def}\\
\dot{Y} & =K\{(N_{L}-1)Y^{\tau}-(N-1)Y+N_{O}X^{\tau}\}\nonumber \\
 & -LY+\Xi_{L}(t),\label{eq:dYdt_GTD_def}\\
F & =K\left\{ (N_{O}-1)X^{\tau}-(N-1)X+N_{L}Y^{\tau}\right\} ,\label{eq:F_GTD_def}
\end{align}
where we have applied a parallel translation so that $\mu_{X}=\mu_{Y}=\mu_{F}=0$
holds for the steady state. Since the system is the steady state under
the synchronization assumption, we can use the Fourier transform to
calculate the covariance matrix \cite{Risken:1989:FPEBook,Guillouzic:1999:SDDE}.
Let $\mathcal{F}[\cdots]$ be the Fourier transform operator, and
let $\nu$ be the Fourier variable. We define $\mathcal{S}_{F}(\nu)=\mathcal{F}\left[\left\langle F(t)F(0)\right\rangle \right]$,
$\mathcal{S}_{X}(\nu)=\mathcal{F}\left[\left\langle X(t)X(0)\right\rangle \right]$,
and $\mathcal{S}_{XF}(\nu)=\mathcal{F}\left[\left\langle X(t)F(0)\right\rangle \right]$.
By virtue of the Wiener--Khinchin theorem, $\mathcal{V}_{X}$, $\mathcal{V}_{F}$,
and $\mathcal{C}_{XF}$ can be obtained by $\mathcal{V}_{X}=(2\pi)^{-1}\int_{-\infty}^{\infty}\mathcal{S}_{X}(\nu)d\nu$,
$\mathcal{V}_{F}=(2\pi)^{-1}\int_{-\infty}^{\infty}\mathcal{S}_{F}(\nu)d\nu$,
and $\mathcal{C}_{XF}=(2\pi)^{-1}\int_{-\infty}^{\infty}\mathcal{S}_{XF}(\nu)d\nu$
(Appendix~\ref{sec:app_glob_time_delay}). Since the integration
is complicated, we evaluate these integrals numerically.

Figure~\ref{fig:delayed_fig}(a) shows $\mathcal{V}_{X}$ and $\mathcal{V}_{X}^{\mathrm{LB}}$
as a function of $R$ for time delay $\tau=0.02$ and $K=1$. The
other parameters are the same as in Fig.~\ref{fig:variance_plot}(a).
In Fig.~\ref{fig:delayed_fig}(a), the dotted and dot-dashed lines
denote $\mathcal{V}_{X}$ and $\mathcal{V}_{X}^{\mathrm{LB}}$ for
$\tau=0.02$, respectively. The solid and long-dashed lines represent
$\mathcal{V}_{X}^{\mathrm{LB}}$ and $\mathcal{V}_{X}$ for the non-delayed
case and are identical to those in Fig.~\ref{fig:variance_plot}(a)
(shown for comparison). We also carried out Monte Carlo simulations
for $\tau=0.02$, and these data are shown by circles and triangles,
which correspond to $\mathcal{V}_{X}$ and $\mathcal{V}_{X}^{\mathrm{LB}}$,
respectively. We again confirm that the inequality of Eq.~\eqref{eq:LB_def}
also holds for the delayed case. Figure~\ref{fig:delayed_fig}(a)
shows that $\mathcal{V}_{X}$ for $\tau=0.02$ is lower than that
of the non-delayed case. Furthermore, $\mathcal{V}_{X}^{\mathrm{LB}}$
is also lower for the delayed case. In order to clarify the effect
of the time delay, we plot $\mathcal{V}_{X}$ and $\mathcal{V}_{X}^{\mathrm{LB}}$
as a function of $\tau$ for $N=N_{O}=10$ and $K=1$ in Fig.~\ref{fig:delayed_fig}(b).
The solid and dashed lines denote $\mathcal{V}_{X}$ and $\mathcal{V}_{X}^{\mathrm{LB}}$,
respectively, obtained analytically, and the circles and triangles
denote $\mathcal{V}_{X}$ and $\mathcal{V}_{X}^{\mathrm{LB}}$, respectively,
obtained by Monte Carlo simulations. For $\tau<0.03$, the analytical
and Monte Carlo results agree, whereas they do not agree for $\tau>0.03$.
This is because $\phi_{j}^{\tau}-\phi_{i}$ in the coupling deviates
from $0$ when $\tau$ increases, which degrades the reliability of
the Taylor approximation. Intriguingly, Fig.~\ref{fig:delayed_fig}(b)
shows that when $\tau$ increases to $\tau\sim0.04$, both $\mathcal{V}_{X}$
and $\mathcal{V}_{X}^{\mathrm{LB}}$ decrease. Since the lower bound
$\mathcal{V}_{X}^{\mathrm{LB}}$ is smaller in the presence of a time
delay, the precision improvement can be ascribed to the increase in
the information flow $\dot{\mathcal{I}}_{X}(X;F)$. Although the time
delay in the Kuramoto model is often studied from the viewpoint of
coherence and incoherence \cite{Yeung:1999:DelayedKuramoto}, we show
that the time delay is beneficial to temporal precision of oscillators.

\section{Conclusion}

We obtained the inequality relating the temporal variance in the Kuramoto
model with the information flow conferred by coupling between oscillators.
CEP is a universal phenomenon and is not limited to coupled oscillators.
For instance, the precision of cellular concentration inference is
substantially improved when the concentration is measured by a population
of cells \cite{Fancher:2017:Sensing}. A similar calculation can be
applied in such cases, which will be left for future studies. 

\appendix

\section{Globally coupled model\label{sec:app_glob_couple}}

We calculate $\mathcal{V}_{X}^{\mathrm{LB}}$ and $\mathcal{V}_{X}$
for a uniformly globally coupled model, where $K_{ij}=K$. The Langevin
equations for $X$ and $Y$ are given by Eqs.~\eqref{eq:dXdt_def_1}
and \eqref{eq:dYdt_def}, respectively. The coupling variable $F$
is 
\begin{equation}
F=KN_{L}(Y-X).\label{eq:glob_F_def}
\end{equation}
Introducing hypothetical delay $h$ in Eq.~\eqref{eq:glob_F_def},
we have 
\begin{equation}
F(t)=KN_{L}(Y(t-h)-X(t-h)).\label{eq:Ft_hyp_delay}
\end{equation}
Assuming that $h$ is sufficiently small, the coupling variable obeys
Eq.~\eqref{eq:dFdt_gl_def}. Equations~\eqref{eq:dXdt_def_1}, \eqref{eq:dYdt_def},
and \eqref{eq:dFdt_gl_def} constitute coupled Langevin equations.
The corresponding FPE is\begin{widetext} 
\begin{align}
\frac{\partial}{\partial t}P(X,Y,F) & =-\frac{\partial}{\partial X}\left\{ -L(X-c)+F\right\} P(X,Y,F)+\frac{D}{N_{O}}\frac{\partial^{2}}{\partial X^{2}}P(X,Y,F)\nonumber \\
 & -\frac{\partial}{\partial Y}\left\{ -L(Y-c)-\frac{F}{R}\right\} P(X,Y,F)+\frac{D}{N_{L}}\frac{\partial^{2}}{\partial Y^{2}}P(X,Y,F)\nonumber \\
 & -\frac{\partial}{\partial F}\frac{1}{h}\left\{ -F+KN_{L}(Y-X)\right\} P(X,Y,F).\label{eq:glob_FPE_def}
\end{align}
\end{widetext}The steady-state distribution of Eq.~\eqref{eq:glob_FPE_def}
is a Gaussian distribution: 
\[
P(X,Y,F)=\mathcal{N}\left(X,Y,F|\boldsymbol{\mu}_{XYF},\boldsymbol{\Sigma}_{XYF}\right),
\]
where $\mathcal{N}(\cdot)$ denotes a multivariate Gaussian distribution,
and $\boldsymbol{\mu}_{XYF}$ and $\boldsymbol{\Sigma}_{XYF}$ are
its mean vector and covariance matrix, respectively. We can obtain
the mean and covariance in the steady state. For $h\rightarrow0$,
the hypothetical delay vanishes, and $F$ reduces to Eq.~\eqref{eq:glob_F_def}.
The mean vector is $\boldsymbol{\mu}_{XYF}=[c,c,0]$, and the covariance
matrix $\boldsymbol{\Sigma}_{XYF}$ is\begin{widetext} 
\begin{equation}
\boldsymbol{\Sigma}_{XYF}=\left[\begin{array}{ccc}
{\displaystyle \frac{\left(KN_{O}+L\right)D}{N_{O}L\left(K\left(R+1\right)N_{O}+L\right)}} & {\displaystyle \frac{DK}{L\left(K\left(R+1\right)N_{O}+L\right)}} & {\displaystyle -\frac{KRD}{K\left(R+1\right)N_{O}+L}}\\
{\displaystyle \frac{DK}{L\left(K\left(R+1\right)N_{O}+L\right)}} & {\displaystyle \frac{\left(KN_{O}R+L\right)D}{RN_{O}L\left(K\left(R+1\right)N_{O}+L\right)}} & {\displaystyle \frac{DK}{K\left(R+1\right)N_{O}+L}}\\
{\displaystyle -\frac{KRD}{K\left(R+1\right)N_{O}+L}} & {\displaystyle \frac{DK}{K\left(R+1\right)N_{O}+L}} & {\displaystyle \frac{N_{O}\left(R+1\right)RK^{2}D}{K\left(R+1\right)N_{O}+L}}
\end{array}\right].\label{eq:cov_Sigma_def}
\end{equation}
\end{widetext}With Eq.~\eqref{eq:I_X_def}, the information flow
is obtained as follows: 
\begin{equation}
\dot{\mathcal{I}}_{X}(X;F)=-\frac{\mathcal{C}_{XF}\left\{ \mathcal{C}_{XF}(D+N_{O}\mathcal{C}_{XF})-N_{O}\mathcal{V}_{F}\mathcal{V}_{X}\right\} }{N_{O}\mathcal{V}_{X}(\mathcal{C}_{XF}^{2}-\mathcal{V}_{F}\mathcal{V}_{X})}.\label{eq:IX}
\end{equation}
From Eqs.~\eqref{eq:cov_Sigma_def} and \eqref{eq:IX}, we obtain
Eq.~\eqref{eq:IX_global}.

\section{Time-delayed model\label{sec:app_time_delay}}

In the main text, we consider the time-delayed case. Assuming that
$\Omega\tau$ is sufficiently small, we can apply the Taylor approximation
to obtain
\begin{align}
\frac{dX}{dt} & =\omega-\Omega-LX+F+\Xi_{O}(t),\label{eq:dXdt_delay_def}\\
\frac{dx_{i}}{dt} & =\omega-\Omega-Lx_{i}+\sum_{j\in V,j\ne i}K_{ij}(x_{j}^{\tau}-x_{i}-\Omega\tau)\nonumber \\
 & +\xi_{i}(t)\hspace*{1em}(i\in V),\label{eq:dxidt_delay_def}\\
\frac{dF}{dt} & =\frac{1}{h}\left\{ -F+\frac{1}{N_{O}}\sum_{i\in V_{O}}\sum_{j\in V,j\ne i}K_{ij}(x_{j}^{\tau}-x_{i}-\Omega\tau)\right\} .\label{eq:dFdt_delay_def}
\end{align}
When deriving Eq.~\eqref{eq:dFdt_delay_def}, we again introduced
the hypothetical time-delay $h$ as in Eq.~\eqref{eq:Ft_hyp_delay}.
Suppose that the mean of $\boldsymbol{x}$ in the steady state is
$[\mu_{x_{1}},\mu_{x_{2}},...,\mu_{x_{N}}]=[c_{x_{1}},c_{x_{2}},...,c_{x_{N}}]$,
where $[c_{x_{1}},c_{x_{2}},...,c_{x_{N}}]$ satisfies the following
relation:
\[
\omega-\Omega-Lc_{x_{i}}+\sum_{j\in V,j\ne i}K_{ij}(c_{x_{j}}-c_{x_{i}}-\Omega\tau)=0\hspace*{1em}(i\in V).
\]
 Equations~\eqref{eq:dXdt_delay_def}--\eqref{eq:dFdt_delay_def}
can be written as
\begin{align}
\frac{dX}{dt} & =-L\left(X-c_{X}\right)+F-c_{F}+\Xi_{O}(t),\label{eq:dXdt_delay_def2}\\
\frac{dx_{i}}{dt} & =-L\left(x_{i}-c_{x_{i}}\right)+\sum_{j\in V,j\ne i}K_{ij}(x_{j}^{\tau}-x_{i}-\Omega\tau)-c_{f_{i}}\nonumber \\
 & +\xi_{i}(t)\hspace*{1em}(i\in V),\label{eq:dxidt_delay_def2}\\
\frac{dF}{dt} & =\frac{1}{h}\left\{ -F+\frac{1}{N_{O}}\sum_{i\in V_{O}}\sum_{j\in V,j\ne i}K_{ij}(x_{j}^{\tau}-x_{i}-\Omega\tau)\right\} ,\label{eq:dFdt_delay_def2}
\end{align}
where 
\begin{align*}
c_{f_{i}} & =\sum_{j\in V,j\ne i}K_{ij}(c_{x_{j}}-c_{x_{i}}-\Omega\tau),\\
c_{X} & =\frac{1}{N_{O}}\sum_{i\in V_{O}}c_{x_{i}},\\
c_{F} & =\frac{1}{N_{O}}\sum_{i\in V_{O}}c_{f_{i}}.
\end{align*}
Following Ref.~\cite{Guillouzic:1999:SDDE}, we can obtain the time
evolution of $P(X,\boldsymbol{x},F)=\int P(X,\boldsymbol{x},F;\boldsymbol{x}^{\tau})d\boldsymbol{x}^{\tau}$
where $P(X,\boldsymbol{x},F;\boldsymbol{x}^{\tau})$ is the probability
density of $X$, $\boldsymbol{x}$, and $F$ at time $t$ and of $\boldsymbol{x}^{\tau}$
at time $t-\tau$. Let $\mathcal{A}(X,\boldsymbol{x},F)$ be an arbitrary
function of $X$, $\boldsymbol{x}$, and $F$. The derivation in Ref.~\cite{Guillouzic:1999:SDDE}
is constructed on the Ito interpretation. If Langevin equations of
interest obey the Stratonovich interpretation, they have to be converted
to equivalent Ito equations \cite{Guillouzic:1999:SDDE}. Since Eqs.~\eqref{eq:dXdt_delay_def2}--\eqref{eq:dFdt_delay_def2}
include only additive noise terms, the calculations afterwards do
not depend on the stochastic integral. Taking the expectation for
$d\mathcal{A}(X,\boldsymbol{x},F)/dt$, we apply the Ito calculus
to obtain\begin{widetext} 
\begin{align*}
\left\langle \frac{d\mathcal{A}}{dt}\right\rangle  & =\left\langle \frac{\partial\mathcal{A}}{\partial X}[-L(X-c_{X})+F-c_{F}]+\frac{1}{h}\frac{\partial\mathcal{A}}{\partial F}\left[-F+\frac{1}{N_{O}}\sum_{i\in V_{O}}\sum_{j\in V,j\ne i}K_{ij}(x_{j}^{\tau}-x_{i}-\Omega\tau)\right]\right.\\
 & +\sum_{i\in V}\frac{\partial\mathcal{A}}{\partial x_{i}}\left[-L(x_{i}-c_{x_{i}})+\sum_{j\in V,j\ne i}K_{ij}(x_{j}^{\tau}-x_{i}-\Omega\tau)-c_{f_{i}}\right]\\
 & \left.+\frac{D}{N_{O}}\frac{\partial^{2}\mathcal{A}}{\partial X^{2}}+\sum_{i\in V}D\frac{\partial^{2}\mathcal{A}}{\partial x_{i}^{2}}+\sum_{i\in V_{O}}\frac{2D}{N_{O}}\frac{\partial^{2}\mathcal{A}}{\partial X\partial x_{i}}\right\rangle _{P(X,\boldsymbol{x},F;\boldsymbol{x}^{\tau})}\\
 & =\left\langle \frac{\partial\mathcal{A}}{\partial X}[-L(X-c_{X})+F-c_{F}]+\frac{1}{h}\frac{\partial\mathcal{A}}{\partial F}\left[-F+\left\langle \frac{1}{N_{O}}\sum_{i\in V_{O}}\sum_{j\in V,j\ne i}K_{ij}(x_{j}^{\tau}-x_{i}-\Omega\tau)\right\rangle _{P(\boldsymbol{x}^{\tau}|X,\boldsymbol{x},F)}\right]\right.\\
 & +\sum_{i\in V}\frac{\partial\mathcal{A}}{\partial x_{i}}\left[-L(x_{i}-c_{x_{i}})+\left\langle \sum_{j\in V,j\ne i}K_{ij}(x_{j}^{\tau}-x_{i}-\Omega\tau)\right\rangle _{P(\boldsymbol{x}^{\tau}|X,\boldsymbol{x},F)}-c_{f_{i}}\right]\\
 & \left.+\frac{D}{N_{O}}\frac{\partial^{2}\mathcal{A}}{\partial X^{2}}+\sum_{i\in V}D\frac{\partial^{2}\mathcal{A}}{\partial x_{i}^{2}}+\sum_{i\in V_{O}}\frac{2D}{N_{O}}\frac{\partial^{2}\mathcal{A}}{\partial X\partial x_{i}}\right\rangle _{P(X,\boldsymbol{x},F)}.
\end{align*}
\end{widetext}Here, we explicitly write the probability density with
which the expectation is calculated. Letting $P(B)$ be a probability
density function of arbitrary random variable(s) $B$ ($\int P(B)dB=1$),
we define $\left\langle \cdots\right\rangle _{P(B)}=\int\cdots P(B)dB$
(e.g., $\left\langle \cdots\right\rangle _{P(\boldsymbol{x}^{\tau}|X,F,\boldsymbol{x})}=\int\cdots P(\boldsymbol{x}^{\tau}|X,\boldsymbol{x},F)d\boldsymbol{x}^{\tau}$).
Since $\mathcal{A}(X,\boldsymbol{x},F)$ is an arbitrary function,
using the integration by parts, we obtain 
\begin{equation}
\frac{\partial}{\partial t}P(X,\boldsymbol{x},F)=\mathbb{L}_{\tau}(X,\boldsymbol{x},F)P(X,\boldsymbol{x},F),\label{eq:FPE_delay_def}
\end{equation}
where $\mathbb{L}_{\tau}(X,\boldsymbol{x},F)$ is the following FPE
operator:\begin{widetext} 
\begin{align}
\mathbb{L}_{\tau}(X,\boldsymbol{x},F) & =-\frac{\partial}{\partial X}[-L(X-c_{X})+F-c_{F}]\nonumber \\
 & -\sum_{i\in V}\frac{\partial}{\partial x_{i}}\left[-L(x_{i}-c_{x_{i}})+\left\langle \sum_{j\in V,j\ne i}K_{ij}(x_{j}^{\tau}-x_{i}-\Omega\tau)\right\rangle _{P(\boldsymbol{x}^{\tau}|X,\boldsymbol{x},F)}-c_{f_{i}}\right]\nonumber \\
 & -\frac{\partial}{\partial F}\frac{1}{h}\left[-F+\left\langle \frac{1}{N_{O}}\sum_{i\in V_{O}}\sum_{j\in V,j\ne i}K_{ij}(x_{j}^{\tau}-x_{i}-\Omega\tau)\right\rangle _{P(\boldsymbol{x}^{\tau}|X,\boldsymbol{x},F)}\right]\nonumber \\
 & +\frac{D}{N_{O}}\frac{\partial^{2}}{\partial X^{2}}+\sum_{i\in V}D\frac{\partial^{2}}{\partial x_{i}^{2}}+\sum_{i\in V_{O}}\frac{2D}{N_{O}}\frac{\partial^{2}}{\partial X\partial x_{i}}.\label{eq:L_tau_def}
\end{align}
\end{widetext}For details of the derivations, please see Ref.~\cite{Guillouzic:1999:SDDE}.
Integrating Eqs.~\eqref{eq:FPE_delay_def} and \eqref{eq:L_tau_def}
with respect to $\boldsymbol{x}$, we obtain
\begin{align}
\frac{\partial P(X,F)}{\partial t} & =-\frac{\partial}{\partial X}[-L(X-c_{X})+F-c_{F}]P(X,F)\nonumber \\
 & +\frac{D}{N_{O}}\frac{\partial^{2}}{\partial X^{2}}P(X,F)\nonumber \\
 & -\frac{\partial}{\partial F}\frac{1}{h}\left\{ -F+\mathcal{H}_{\tau}(X,F)\right\} P(X,F),\label{eq:PXF_delay}
\end{align}
where 
\[
\mathcal{H}_{\tau}(X,F)=\left\langle \frac{1}{N_{O}}\sum_{i\in V_{O}}\sum_{j\in V,j\ne i}K_{ij}(x_{j}^{\tau}-x_{i}-\Omega\tau)\right\rangle _{P(\boldsymbol{x},\boldsymbol{x}^{\tau}|X,F)}.
\]
Since Eq.~\eqref{eq:PXF_delay} is identical to Eqs.~\eqref{eq:FPE_XF}--\eqref{eq:JF_def}
when replacing $\mathcal{H}(X,F)$ with $\mathcal{H}_{\tau}(X,F)$
and $F\rightarrow F-c_{F}$ in a drift term, we obtain
\[
\left\langle (X-c_{X})^{2}\right\rangle \ge\frac{D}{N_{O}L^{2}}\left(L-\dot{\mathcal{I}}_{X}(X;F)\right),
\]
which is the same inequality as the non-delayed model (Eq.~\eqref{eq:LB_def}).

\section{Globally coupled time-delay model\label{sec:app_glob_time_delay}}

We consider a globally coupled case $K_{ij}=K$ to calculate $\mathcal{V}_{X}^{\mathrm{LB}}$
and $\mathcal{V}_{X}$. The calculations can be classified into two
cases: $N_{L}\ge1$ and $N_{L}=0$. 

For $N_{L}\ge1$, we obtain Eqs.~\eqref{eq:dXdt_GTD_def}--\eqref{eq:F_GTD_def}
for $X$, $Y$, and $F$. We need to calculate the covariance matrix
of $X$ and $F$. These quantities can be obtained via the Fourier
transform \cite{Guillouzic:1999:SDDE}. The detailed procedures (known
as the Rice method) are shown in Ref.~\cite{Risken:1989:FPEBook}.
For an arbitrary random variable $A(t)$, we can define the Fourier
transform and its inverse as follows: 
\begin{align*}
\widetilde{A}(\nu) & =\mathcal{F}\left[A(t)\right]=\int_{-\infty}^{\infty}e^{-\mathrm{i}\nu t}A(t)dt,\\
A(t) & =\mathcal{F}^{-1}\left[\widetilde{A}(\nu)\right]=\frac{1}{2\pi}\int_{-\infty}^{\infty}e^{\mathrm{i}\nu t}\widetilde{A}(\nu)d\nu,
\end{align*}
where $\mathrm{i}$ in a roman typeface denotes the imaginary unit.
Then, applying the Fourier transform, we obtain 
\begin{equation}
\left[\begin{array}{c}
\widetilde{X}(\nu)\\
\widetilde{Y}(\nu)\\
\widetilde{F}(\nu)
\end{array}\right]=\mathcal{M}\left[\begin{array}{c}
\widetilde{\Xi}_{O}(\nu)\\
\widetilde{\Xi}_{L}(\nu)\\
0
\end{array}\right],\label{eq:XYF_Fourier_equation}
\end{equation}
where $\mathcal{M}$ is a regular matrix defined by\begin{widetext}
\[
\mathcal{M}=\left[\begin{array}{ccc}
L+\mathrm{i}\nu & 0 & -1\\
-e^{-\mathrm{i}\nu\tau}KN_{O} & L+K\left\{ -1-e^{-\mathrm{i}\nu\tau}(N_{L}-1)+N_{L}+N_{O}\right\} +\mathrm{i}\nu & 0\\
K\left\{ -1+N_{L}-e^{-\mathrm{i}\nu\tau}(N_{O}-1)+N_{O}\right\}  & -e^{-\mathrm{i}\nu\tau}KN_{L} & 1
\end{array}\right]^{-1}.
\]
\end{widetext}We can consider the Fourier transform of the correlation
function: $\mathcal{S}_{X}(\nu)=\mathcal{F}\left[\left\langle X(t)X(0)\right\rangle \right]$,
$\mathcal{S}_{F}(\nu)=\mathcal{F}\left[\left\langle F(t)F(0)\right\rangle \right]$,
$\mathcal{S}_{XF}(\nu)=\mathcal{F}\left[\left\langle X(t)F(0)\right\rangle \right]$,
$\mathcal{S}_{O}(\nu)=\mathcal{F}\left[\left\langle \Xi_{O}(t)\Xi_{O}(0)\right\rangle \right]=2D/N_{O}$,
and $\mathcal{S}_{L}(\nu)=\mathcal{F}\left[\left\langle \Xi_{L}(t)\Xi_{L}(0)\right\rangle \right]=2D/N_{L}$.
Let $A$ and $B$ be arbitrary random variables. According to the
Wiener--Khinchin theorem, the following relation holds \cite{Risken:1989:FPEBook}:
\begin{align}
\left\langle \widetilde{A}(\nu)\widetilde{B}^{*}(\nu^{\prime})\right\rangle  & =2\pi\delta(\nu-\nu^{\prime})\mathcal{S}_{AB}(\nu),\label{eq:WH1}\\
\mathcal{S}_{AB}(\nu) & =\mathcal{F}\left[\left\langle A(t)B^{*}(0)\right\rangle \right],\label{eq:WH2}
\end{align}
where superscript $*$ denotes the complex conjugate. From Eqs.~\eqref{eq:XYF_Fourier_equation}--\eqref{eq:WH2},
we obtain 
\begin{align}
\mathcal{S}_{X}(\nu) & =\left|\mathcal{M}_{11}\right|^{2}\mathcal{S}_{O}(\nu)+\left|\mathcal{M}_{12}\right|^{2}\mathcal{S}_{L}(\nu),\label{eq:SX1}\\
\mathcal{S}_{F}(\nu) & =\left|\mathcal{M}_{31}\right|^{2}\mathcal{S}_{O}(\nu)+\left|\mathcal{M}_{32}\right|^{2}\mathcal{S}_{L}(\nu),\label{eq:SF1}\\
\mathcal{S}_{XF}(\nu) & =\mathcal{M}_{11}\mathcal{M}_{31}^{*}\mathcal{S}_{O}(\nu)+\mathcal{M}_{12}\mathcal{M}_{32}^{*}\mathcal{S}_{L}(\nu),\label{eq:SXF}
\end{align}
where $\mathcal{M}_{ij}$ is the $i,j$-th element of $\mathcal{M}$.
When we assume a Gaussian distribution for $P(X,F)$, from Eq.~\eqref{eq:IX},
it is sufficient to calculate $\mathcal{V}_{X}$, $\mathcal{V}_{F}$,
and $\mathcal{C}_{XF}$ for information flow $\dot{\mathcal{I}}_{X}$.
From Eq.~\eqref{eq:WH2}, the inverse Fourier transform yields variance
$\mathcal{V}_{X}$ and $\mathcal{V}_{F}$ and covariance $\mathcal{C}_{XF}$:
\begin{align}
\mathcal{V}_{X} & =\left\langle X(t)^{2}\right\rangle =\frac{1}{2\pi}\int_{-\infty}^{\infty}\mathcal{S}_{X}(\nu)d\nu,\label{eq:VX}\\
\mathcal{V}_{F} & =\left\langle F(t)^{2}\right\rangle =\frac{1}{2\pi}\int_{-\infty}^{\infty}\mathcal{S}_{F}(\nu)d\nu,\label{eq:VF}\\
\mathcal{C}_{XF} & =\left\langle X(t)F(t)\right\rangle =\frac{1}{2\pi}\int_{-\infty}^{\infty}\mathcal{S}_{XF}(\nu)d\nu.\label{eq:VXF}
\end{align}
Since it is unlikely that we can obtain the inverse Fourier transforms
of Eqs.~\eqref{eq:VX}--\eqref{eq:VXF} analytically, we calculated
the transforms numerically.

We next consider the case in which $N_{L}=0$. In a similar way, we
have the following coupled equations: 
\begin{align*}
\frac{dX}{dt} & =-LX+F+\Xi_{O}(t),\\
F & =K(N-1)(X^{\tau}-X).
\end{align*}
The Fourier transform yields 
\begin{equation}
\left[\begin{array}{c}
\widetilde{X}(\nu)\\
\widetilde{F}(\nu)
\end{array}\right]=\mathcal{L}\left[\begin{array}{c}
\widetilde{\Xi}_{O}(\nu)\\
0
\end{array}\right],\label{eq:XYF_Fourier_equation2}
\end{equation}
where $\mathcal{L}$ is a regular matrix defined as 
\[
\mathcal{L}=\left[\begin{array}{cc}
L+\mathrm{i}\nu & -1\\
K(N-1)(1-e^{-\mathrm{i}\nu\tau}) & 1
\end{array}\right]^{-1}.
\]
Then, the Fourier transforms of the correlation functions are 
\begin{align}
\mathcal{S}_{X}(\nu) & =\left|\mathcal{L}_{11}\right|^{2}\mathcal{S}_{O}(\nu),\label{eq:SX1b}\\
\mathcal{S}_{F}(\nu) & =\left|\mathcal{L}_{21}\right|^{2}\mathcal{S}_{O}(\nu),\label{eq:SF1b}\\
\mathcal{S}_{XF}(\nu) & =\mathcal{L}_{11}\mathcal{L}_{21}^{*}\mathcal{S}_{O}(\nu).\label{eq:SXFb}
\end{align}
Again, the variance and covariance can be obtained by the inverse
Fourier transform of Eqs.~\eqref{eq:VX}--\eqref{eq:VXF}.

\section{Monte Carlo simulation\label{sec:app_monte_carlo}}

In order to solve Langevin equations numerically, we use the Euler
method shown in Ref.~\cite{Risken:1989:FPEBook} with time step $\epsilon=0.001$.
When calculating the information flow $\dot{\mathcal{I}}_{X}(X;F)$,
we use the following approximation from Eq.~\eqref{eq:I_X_def}:
\begin{equation}
\dot{\mathcal{I}}_{X}(X;F)\simeq-\frac{1}{\epsilon}\left\langle \ln\left(\frac{P(F(t)|X(t+\epsilon))}{P(F(t)|X(t))}\right)\right\rangle .\label{eq:IX_approx_def}
\end{equation}
Here, we use $F(t)$ based on Eq.~\eqref{eq:F_def_Kuramoto} {[}non-delayed
case{]} or Eq.~\eqref{eq:fi_def_delay} {[}time-delayed case{]}.
In Eq.~\eqref{eq:IX_approx_def}, we need to calculate the probability
density $P(X,F)$. Assuming a multivariate Gaussian distribution for
$P(X,F)$, we empirically calculate the mean vector and covariance
matrix for $P(X,F)$. Then, we can numerically calculate the information
flow $\dot{\mathcal{I}}_{X}(X;F)$ as the average of $2\times10^{6}$
samples. Regarding the number of oscillators, we use $N_{O}=10$ and
$N_{L}=0$--$10$ ($0\le R\le1$). Note that when we employ an excessively
large $N$, the simulation suffers from an artifact due to the time
discretization, because discretized coupling terms $f_{i}(t)\epsilon$
are on the order of $O(N\epsilon)$.

\end{document}